\documentclass[aps,prd,onecolumn,superscriptaddress,showpacs,floatfix]{revtex4}
\usepackage{bm, natbib, hyperref, amsmath}
\usepackage{graphicx}

\newcommand{\beq}{\begin{equation}}
\newcommand{\eeq}{\end{equation}}
\newcommand{\beqa}{\begin{eqnarray}}
\newcommand{\eeqa}{\end{eqnarray}}

\def\apj{Astrophys.\ J.\ }
\def\apjl{Astrophys.\ J.\ Lett.\ }
\def\apjs{Astrophys.\ J.\ Suppl.\ Ser.\ }
\def\mnras{Mon.\ Not.\ R.\ Astron.\ Soc.\ }

\def\prl{Phys.\ Rev.\ Lett.\ }
\def\prd{Phys.\ Rev.\ D\ }

\def\plb{Phys.\ Lett.\ B\ }
\def\physrep{Phys.\ Rep.\ }
\def\jcap{J.\ Cosmol.\ Astropart.\ Phys.\ }
\def\grg{Gen.\ Relativ.\ Gravit.\ }
\def\cqg{Class.\ Quant.\ Grav.\ }

\begin{document}

\title{Tests of Gravity from Imaging and Spectroscopic Surveys}
\author{Jacek Guzik}
\email{guzikj@gmail.com}
\affiliation{Department of Physics and Astronomy, University of Pennsylvania, Philadelphia, PA 19104, U.S.A.}
\affiliation{Astronomical Observatory, Jagiellonian University, Orla 171, 30-244 Krak\'ow, Poland}
\author{Bhuvnesh Jain}
\email{bjain@physics.upenn.edu}
\affiliation{Department of Physics and Astronomy, University of Pennsylvania, Philadelphia, PA 19104, U.S.A.}
\author{Masahiro Takada}
\email{masahiro.takada@ipmu.jp}
\affiliation{Institute for the Physics and Mathematics of the Universe (IPMU), The University of Tokyo,
Chiba 277-8582, Japan}

\begin{abstract}
Tests of gravity on large-scales in the universe can be made using
both imaging and spectroscopic surveys. The former allow for
measurements of weak lensing, galaxy clustering and cross-correlations
such as the ISW effect. The latter probe galaxy dynamics through 
redshift space distortions. We use a set of basic observables, 
namely lensing power spectra,
galaxy-lensing  and galaxy-velocity cross-spectra in multiple
redshift bins (including their covariances), to estimate the ability 
of upcoming surveys to test
gravity theories. We use a two-parameter description of gravity that
allows for the Poisson equation and the ratio of metric potentials to
depart from general relativity. We find that the combination of
imaging and spectroscopic observables is essential in making 
robust tests of gravity theories. The range of scales and redshifts 
best probed by upcoming surveys is discussed. We also compare our 
parametrization to others used in the literature, in particular the 
$\gamma$ parameter modification of the growth factor. 
\end{abstract}


\maketitle

\section{Introduction}

General relativity (GR) plus the Standard Model of particle
physics can only account for about $4\%$ of the energy density inferred from
observations.  By introducing dark matter and dark  
energy, which account for the remaining $96\%$ of the total 
energy budget of the universe,  cosmologists have been able to
account for a wide range of  observations, from  
the overall expansion of the universe to various measures of 
large scale structure~\cite{Reviews}. 

The dark matter/dark energy scenario assumes the validity of GR at
galactic and cosmological scales and introduces exotic components of
matter and energy to account for observations. Since 
GR has not been tested independently on these scales, a natural
alternative is that GR itself needs to be modified on large scales. 
This possibility, that modifications of the law of gravity
on galactic and cosmological scales can replace dark matter and/or 
dark energy, has become an area of active research in recent years. 
Attempts have been made to modify GR with a focus on 
galactic~\cite{MOND} or cosmological
scales~\cite{DGP,fR,Sahni2005}.  The DGP model~\cite{DGP}, in which
gravity lives in a 5-dimensional space-time, can produce a late time 
acceleration of the universe. Adding a correction term $f(R)$ to the
Einstein-Hilbert action \cite{fR} also allows late time acceleration of
the universe   to be realized.  

In this paper we will focus on modified gravity (MG) theories that are
designed as an alternative to dark energy (DE) to produce the present day 
acceleration of the universe. In these models, such as DGP and $f(R)$
models, gravity at late cosmic times and on large-scales departs from
the predictions of GR.  By design, successful MG models
are difficult to distinguish from viable DE models using observations of
the expansion history  of the universe. However, in general they
predict a different growth of perturbations which can be tested using
observations of large-scale structure (LSS) 
\cite{Yukawa,Stabenau2006,Skordis06,Dodelson06,DGPLSS,consistencycheck,Koyama06,
fRLSS,Zhang06,Bean06,linder05, huterer_linder07,Uzan06,Caldwell07,Amendola07,SongHuSaw}.  

LSS in MG theories can be more complicated to predict, but is also 
richer because different observables like lensing and galaxy
clustering probe independent perturbed variables. This differs from 
conventional DE scenarios where the linear growth factor of the
density field fixes all observables on sufficiently large-scales. 
Theories of LSS in these modified gravity models are still in their infancy. 
Most studies have focused on probes of a single
growth factor with one or a few observables.  
Recent predictions for discriminatory power of different observables
could be found in \citep{Zhang7,acquaviva08, schmidt08, zhao08,song_dore08}.  

We study tests of gravity that can be made with 
a combination of imaging and spectroscopic surveys 
Our emphasis will be on
model-independent constraints of MG enabled by combining different
observables. Carrying out robust tests of MG in practice is challenging 
as in the absence of a fundamental theory, the modifications to
gravity are often parametrized by 
free functions, to be fine tuned and fixed by observations. 
Recently the Parametrized Post-Friedman approach has been 
suggested as an attempt to describe a variety of gravity theories 
\citep{HuSaw}.

In \S \ref{sec:method} we describe the ingredients of our modeling - 
parametrization of the MG models (\S \ref{parametrize}), observables 
used for forecasting (\S \ref{sec:observ}) and covariances between
observables (\S \ref{errors}). 
In \S \ref{sec:results} forecasts for upcoming imaging and 
spectroscopic surveys are presented. We conclude in \S \ref{sec:conclusions}.
The derivation of formulas for some of the covariances 
used in the main text are given in the Appendix 
\ref{app:name} and results for 
alternative MG parametrization in Appendix \ref{app:mg}. 

\section{Method}
\label{sec:method}

\subsection{Parametrization of modifications to gravity} 
\label{parametrize}

We are concerned with sub-horizon scales 
that satisfy the
quasi-static, Newtonian approximation. In this regime, two effective
functions characterize the departure of modified gravity
theories with scalar perturbations from general relativity. We neglect
additional fields that are expected to play a role on small, nonlinear
scales to drive the theory to GR. 
Scalar perturbations in a homogeneous and isotropic universe   
can be described in the Newtonian (longitudinal) gauge by two potentials $\Phi(t,{\bf x})$ and $\Psi(t,{\bf x})$
as follows \citep{mfb_92}
\begin{equation}
   ds^2 = - \left[1+2\Psi(t,{\bf x})\right] dt^2 + a^2(t) \left[1-2\Phi(t,{\bf x})\right] \left[d\chi^2 + r(\chi)^2 d\Omega^2\right],
\end{equation}
where $a(t)$ is the scale factor and $r(\chi)$ the comoving angular-diameter distance.
Throughout this paper we assume that the universe is spatially flat, so that
 $r(\chi) = \chi$.  
In GR, neglecting sources of the anisotropic stress in the energy-momentum tensor, the relation $\Phi = \Psi$ holds
\citep{mfb_92}. 
Moreover, the curvature potential $\Phi$ is related to the mass density distribution 
$\rho(a,{\bf x}) = \bar{\rho}(a) \delta(a,{\bf x})$ through 
the Poisson equation, which can be altered in MG theories. 
We assume that the Fourier-space analogue of the Poisson equation becomes 
\begin{equation}
   -k^2 \Phi(a,{\bf k}) = 4 \pi a^2  G g(k) \bar{\rho} \, \delta(a,{\bf k}), 
    \label{eqn:mod_phi}
\end{equation}
where $G g(k)$ is the effective gravitational constant. 
The relation between the curvature potential $\Phi$ and the 
Newtonian potential  $\Psi$ in MG theories is parametrized as
\begin{equation}
	\eta(k) = \frac{\Phi}{\Psi}.  
    \label{eqn:mod_psi}
\end{equation}
There are a variety of parametrization of MG in the literature; the
one described above has been suggested by a number of authors 
\citep{JZ,zhao08,schmidt08}. 
A more general description of modified gravity models via a 
Parametrized-Post-Friedman approach \citep{HuSaw} extends to the 
superhorizon regime, but we opt for the two free parameters described
above as we restrict ourselves to the quasi-static Newtonian regime. 
It is expected to be a generic feature of MG that the linear growth 
of structure becomes scale dependent and its time evolution gets changed with respect to the GR case \citep{HuSaw}. 

The simplest possibility is that the functions $g(k)$ and $\eta(k)$ 
do not depend either on time or on scale, 
hence we denote $g(k)=g_0$ and $\eta(k)=\eta_0$. 
In  GR both parameters are unity, 
which is their fiducial value throughout the paper. 
Other parametrization will be discussed below; in some cases a specific
scale or time dependence allows for easier constraints on modified 
gravity parameters. 

The growth of structure in a CDM dominated universe 
with MG is given by 
\begin{equation}
  \ddot{\delta}(a,k) +2 H(a) \dot{\delta}(a,k) - \frac{g(k)}{\eta(k)} 4 \pi G \bar{\rho} \delta(a,k) = 0, 
  \label{eqn:growth}
\end{equation}
where the expansion history, 
given by the Hubble parameter $H = \dot{a}/a$, will be taken to be 
identical to that in the standard $\Lambda$CDM cosmology. 
The growth of structure, described by the evolution of $\delta(a,k)$ 
as given in Eqn. \ref{eqn:growth}, depends on the
ratio of $g(k)$ and $\eta(k)$.  
The linear growth factor 
$D(a) \equiv \delta(a,k)/\delta(a_i,k)$,  which describes evolution of 
matter density perturbations 
(see Eqn. (\ref{eqn:growth}))
relative to their initial values $a=a_i$, is sensitive to this ratio. 
The relevant growth factor for matter peculiar velocities (in fact, its divergence) relative to the density evolution 
is given by  $f(a) \equiv \frac{d\ln D}{d\ln  a}$. 

We also consider a popular parametrization of MG which is based on the
growth exponent $\gamma$ in the growth rate function $f(a)$ \citep{linder05, huterer_linder07}. 
For GR with a cosmological constant
it can be expressed as $f(a) = \Omega_m(a)^\gamma$, where $\gamma = 0.55$ 
and $\Omega_m(a)$ is the total matter density parameter. 
For the DGP model
the growth exponent is $\gamma = 0.68$ \citep{linder05}. 
Other parametrization are discussed in the Appendix. 

\begin{figure}
\includegraphics[width=17cm]{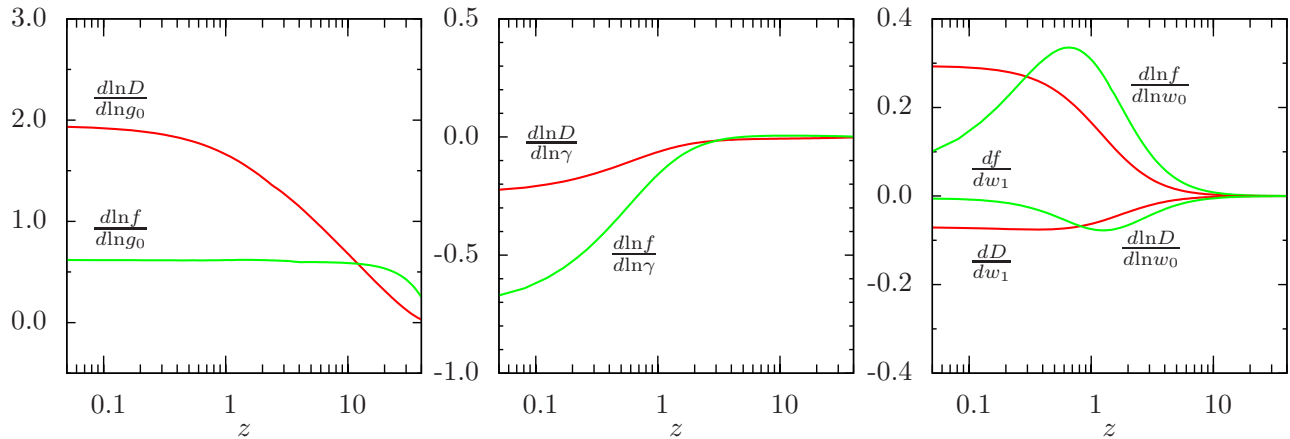}
\caption{
\label{df_sensitivity} 
Sensitivity of the growth function $D(a)$ and the growth rate $f(a)$ 
to the modified gravity parameter $\mu_0$ (left panel), 
growth exponent $\gamma$ (middle panel), 
and the dark energy equation of state parameters $w_0$ and $w_1$ 
(right panel) as a function of redshift. 
}
\end{figure}

The sensitivity of the growth factor $D(a)$ and the growth rate function $f(a)$  to  $\eta_0$ is shown in the Fig. \ref{df_sensitivity}.  
The effect of MG appears to be 
significant as for the redshift range we are interested in ($z<1.5$) 
the change of $D(a)$
is $\sim 2\%$ and $f(a)$ is $\sim 0.5\%$ if 
$g_0$ is changed by $1\%$ (the effect of $\eta_0$ has the same
effect but with opposite sign). If the growth exponent parametrization is 
employed the change in $D(a)$ and $f(a)$ is significantly smaller. 
For comparison we also show the response of 
$D(a)$ and $f(a)$ to the change of the dark energy equation
of state parameters $w_0$ and $w_1$ defined as $w=w_0+ w_1 (1-a)$. 

\subsection{Observables}
\label{sec:observ}

We consider the signal from weak gravitational lensing 
together with galaxy clustering in redshift space.
In metric theories 
of gravity the light deflection angle $\alpha$ 
is given by the transverse gradient of the sum of the  metric potentials:  
$\alpha(\theta) =  \nabla_{\perp} (\Phi+\Psi)$ (see e.g. \cite{carroll}). 
Therefore observed shapes of galaxies and their correlations as 
described by weak lensing power spectra are dependent on both metric 
potentials. The clustering of matter is governed by the growth equation 
(\ref{eqn:growth}) which 
is dependent only upon the Newtonian potential. 
Different combinations of the information from weak lensing and
redshift space galaxy clustering has been used to forecast 
tests of MG models \citep{Zhang7,acquaviva08, song_dore08}. 

For weak lensing we use two observables - the 
correlations between shapes of galaxies (cosmic shear) quantified 
by its power spectrum  $C_{\kappa \kappa}(l)$ and correlations 
between the foreground galaxy distribution and shapes of background 
galaxies (galaxy-galaxy lensing) described by the $C_{g \kappa}(l)$  
cross-power spectrum. 
We use the convergence field $\kappa$ for simplicity as the power
spectra defined below for $\kappa$ are identical to the shear power 
spectra \citep{BS00}. 
It is given by: $\kappa({\bf \theta}) \equiv \frac{1}{2} \nabla_{{\bf
    \theta}} \, \alpha({\bf \theta})$. 
The relation between the convergence and the metric potentials is 
given by the line-of-sight projection:
\begin{equation}
   \kappa({\bf \theta}) = \frac{1}{2} \int_0^{z_s} \frac{dz}{H(z)} \frac{r(z) r(z_s,z)}{r(z_s)} \nabla_{{\bf
    \theta}}^2 (\Phi + \Psi),
   \label{eqn:kappa}
\end{equation}
where $r(z)$ is the comoving angular-diameter distance between observer
and a lens at redshift $z$. We take the 
sources to lie at redshift $z_s$. 

The metric potentials are related to the mass distribution  as given by Eqns. (\ref{eqn:mod_phi}) and (\ref{eqn:mod_psi}), 
so the lensing power spectra can  be expressed in terms of the 
three-dimensional mass power spectrum $P_{\delta \delta}(a,k)$. 
In the small-sky-patch limit (we work with 
scales smaller than $\sim 6^{\circ}$  or $l > 30$), 
the Limber approximation \citep{1953ApJ...117..134L} gives 
\begin{widetext}
\begin{eqnarray}
\label{cls1}
C_{\kappa_i \kappa_j}(l) & = & \frac{9}{4} \Omega_m^2 H_0^4   
  \int_0^{\infty} \frac{dz}{H(z)} \, \frac{1}{a^2(z)} \,  
  \left[ g(k)\frac{1+\eta(k)}{\eta(k)} \right]^2 \,
  P_{\delta \delta}(k,z) W_L(z,z_i) W_L(z,z_j),
\end{eqnarray}
\end{widetext}
where the lensing weight function 
\begin{eqnarray}
  W_L(z,z_k) = \int_{z_k} dz_k \; \frac{dn_b}{dz_k} \, \frac{r(z_k,z)}{r(z_k)},
\end{eqnarray}
depends on the geometry and the redshift 
distribution of lensed galaxies $dn_b/dz$. 
The three-dimensional wavenumber $k$ is given by $k=l/r(z)$. 
We use lensing tomography 
\citep{1999ApJ...522L..21H} by dividing the galaxy distribution 
into $N_z$ bins in redshift. Hence, instead of one projected power 
spectrum we obtain 
$N_z(N_z+1)/2$ power spectra $C_{\kappa_i \kappa_j}(l)$ that carry
additional information about the growth of structure. 
Similarly for the galaxy-shear power spectra we have 
\begin{widetext}
\begin{eqnarray}
  C_{g_i \kappa_j}(l) & = & \frac{3}{2} \Omega_m H_0^2 
  \int_{z_i} dz_i \; \frac{b(z_i)}{a(z_i) \, r(z_i)} \, \frac{dn_f}{dz_i} \, 
  \left[ g(k)\frac{1+\eta(k)}{\eta(k)} \right] \,
  P_{\delta \delta}(k,z_i)  W_L(z,z_j). 
\label{cls2} 
\end{eqnarray}
\end{widetext}
We assume that the distribution of galaxies  $\delta_g$  
is a biased tracer of the mass
distribution $\delta$ but their relation is local and given by a 
bias factor $b$ which may
depend on time, $\delta_g(z_i) = b(z_i) \delta(z_i)$. 
A non-zero $C_{g_i \kappa_j}$ is obtained when galaxies $g_i$ are in front of
the source galaxies, which requires $i \leq j$. 
We also compute the galaxy-galaxy 
projected power spectrum which will not be used as an observable 
but is required in making forecasts for MG by means of the Fisher matrix 
approach. It is given by 
\begin{widetext}
\begin{eqnarray}
  C_{g_i g_j}(l) & = &  \delta_{ij} 
  \int_{z_i} dz_i \; \frac{b^2(z_i)}{r^2(z_i)} \, H(z_i) \, \left[\frac{dn_f}{dz_i}\right]^2 \, P_{\delta \delta}(k,z_i),
  \label{cls3}
\end{eqnarray}
\end{widetext}
where we have assumed that galaxies in two redshift bins are not 
correlated with each other, a good approximation
for wide enough redshift bins. 

Note that the `observed' power spectra differ from the spectra 
in Eqns. (\ref{cls1}), (\ref{cls2}), (\ref{cls3})  
because the effect of discrete sampling of the underlying 
convergence and galaxy density fields should be taken into account. 
It leads to shot (shape) noise terms in the case of the `observed' galaxy (shear) power 
spectra 
\begin{eqnarray}
\hat{C}_{g_i g_j}  &=& C_{g_i g_j} +  
\delta_{ij}/n^{\mathrm{2d}}_{\mathrm{g}}, \\
\hat{C}_{\kappa_i \kappa_j}  &=& C_{\kappa_i \kappa_j} + \sigma_e^2 
\delta_{ij}/n^{\mathrm{2d}}_{\mathrm{g}}, \\
\hat{C}_{g_i \kappa_j} &=& C_{g_i \kappa_j},
\end{eqnarray}
where $n^{\mathrm{2d}}_{\mathrm{g}}$ is the projected density of galaxies.
The shape noise term, proportional to $\sigma_e$, accounts for the intrinsic ellipticities of 
galaxies and its value (per component) is taken to be $0.4/\sqrt{2}$. 
The cross-power spectra $C_{g_i \kappa_j}$ are immune to the discrete 
sampling noise. 

Modifications to GR enter the projected power spectra through the 
lensing-specific factor 
$g(k)\frac{1+\eta(k)}{\eta(k)}$ which is 
responsible for the relation between 
the mass distribution and the metric potentials. 
Moreover, the evolution of structure as expressed by growth functions $D(a)$ 
and $f(a)$ is affected by changes in gravity as a result of 
modifications to Eqns. (\ref{eqn:mod_phi}) and (\ref{eqn:mod_psi}). 

In physical space, the power spectrum of the galaxy distribution 
$P_{gg}(k)$ is expected to be isotropic, 
but in redshift space peculiar velocities distort the distribution 
of galaxies along the line of sight. 
The radial component of peculiar velocities cause 
the observable redshift-space power spectrum  $P^{(s)}_{gg}(k, \mu_k)$
to be `squashed' along the line of sight on 
large scales (in the linear regime) and to produce pronounced
`finger-of-God' features on small scales 
(in the nonlinear regime) \citep{1987MNRAS.227....1K, 1998evun.work..185H}. 
The directional dependence of $P^{(s)}_{gg}$ is given by
$\mu_k \equiv k_{\parallel}/k$, 
which depends on the angle between a wave vector ${\bf k}$ and
the line-of-sight direction. 

Although the picture
is more complicated in reality (see \citep{2004PhRvD..70h3007S} 
for a detailed discussion),  it is a good approximation to decompose the
redshift space power spectrum in terms of three isotropic 
power spectra relating the 
galaxy overdensity  $\delta_g$ and peculiar velocities ${\bf v}$:
the galaxy power spectrum $P_{gg}(k)$,
the velocity power spectrum $P_{vv}(k)$ 
and the cross power spectrum $P_{gv}(k)$ as follows 
\citep{1987MNRAS.227....1K, 2004PhRvD..70h3007S}
\begin{eqnarray}
P^{(s)}_{gg}(k, \mu_k) &=& \left[ P_{gg}(k) + 2 \mu_k^2 P_{gv}(k) + \mu_k^4 P_{vv}(k) \right] F(k^2 \mu^2_k \sigma^2_v), 
        \label{psz:decomp}
\end{eqnarray}
where the term $F(k^2 \mu^2_k \sigma^2_v)$ describes non-linear 
velocity dispersion effects. We set $F \approx 1$, which is a valid on
sufficiently large scales for forecasting purposes.
As before, we assume that galaxies are biased tracers 
of mass and the bias is time-dependent but 
scale independent. 
What we refer to as the velocity
is actually the velocity divergence, which is related to the mass 
distribution through the continuity equation
in the linear regime $ \dot{\delta} + \nabla \cdot {\bf v}/a = 0$  (see \citep{2002PhR...367....1B}).  
Even if gravity is modified the continuity equation stays unchanged  as 
long as there are no components interacting with matter. 

The redshift space power spectrum in the form (\ref{psz:decomp})
shows a distinctive pattern in its angular dependence  
which is important in obtaining the component power spectra from $P^{(s)}_{gg}$;  
it has been used to measure
$P_{gg}(k)$, $P_{gv}(k)$ and $P_{vv}(k)$  power spectra
from the 2dF and SDSS galaxy surveys \citep{2002MNRAS.335..887T,2004ApJ...606..702T}.
Moreover, this decomposition is immune to the scale dependence of 
both the galaxy bias and growth functions 
as long as the angular structure is preserved \citep{2004PhRvD..70h3007S}.

The galaxy-velocity power spectrum $P_{gv}$ which we would like to use in constraining MG models
is not a direct observable. We construct an  estimator 
of $P_{gv}$ band-power spectra  as follows 
\begin{equation}
	\hat{P}_{gv}(k_i) =  \frac{1}{N_k} \sum_{k,\mu} W_{gv}(\mu) \hat{P}^{(s)}_{gg}(k,\mu), \label{eqn:pgv}
\end{equation}
where the summation is carried out over modes in a spherical shell of 
radius $k_i$, and the weight function $W_{gv}(\mu)$ is given by
\begin{equation}
	W_{gv}(\mu) = \frac{15}{4}P_2(\mu)-\frac{135}{8}P_4(\mu), \label{eqn:w2}
\end{equation}
in terms of Legendre polynomials $P_l(\mu)$ of order $l$
(see Appendix A for motivation of this expression). 
The summation in Eqn. (\ref{eqn:pgv}) is carried out over  
modes contained  in spherical shells in 
Fourier space which satisfy the following condition
$k_i-\Delta k_i/2\le |\bm{k}|\le k_i+\Delta k_i/2$. 
The volume $V_k$ of this shell  
can be approximated by $V_k= 4 \pi k_i^2 \Delta k_i$. The 
fundamental volume can be expressed in terms of the survey volume $V_s$ 
as $V_F = (2\pi)^3/V_s$. The 
number of modes in the volume $V_k$ is then given by $N_k =V_k/V_F$.  
We assume that the power spectrum $\hat{P}^{(s)}_{gg}(k,\mu)$ 
does not vary significantly in the shell as a function of 
$|\bm{k}|$. We also assume that the survey is 
large enough that the fundamental volume is much smaller than $V_k$.
$\hat{P}_{gv}(k_i)$ is taken as an observable in the Fisher matrix 
analysis. Note that the redshift space power spectrum does not get 
any additional modifying factors in MG
(except for the growth functions), as the
continuity equation on which the relation between $P^{(s)}_{gg}$ 
and $P_{gv}$ is based remains the same. 

In the present work we focus on the galaxy-velocity 
cross power spectrum $P_{gv}(k)$ and the information it can deliver about 
the growth of structure and galaxy bias.
The bias factor appears linearly in our observables, both in projected power spectra as well as in $P_{gv}(k)$,
and is degenerate with the growth functions. However, when information 
in $P_{gv}(k)$ is combined with tomographic measurements of the weak 
lensing signal it should allow for breaking this degeneracy \citep{Zhang7}. 

\begin{figure*}
\includegraphics[width=16cm]{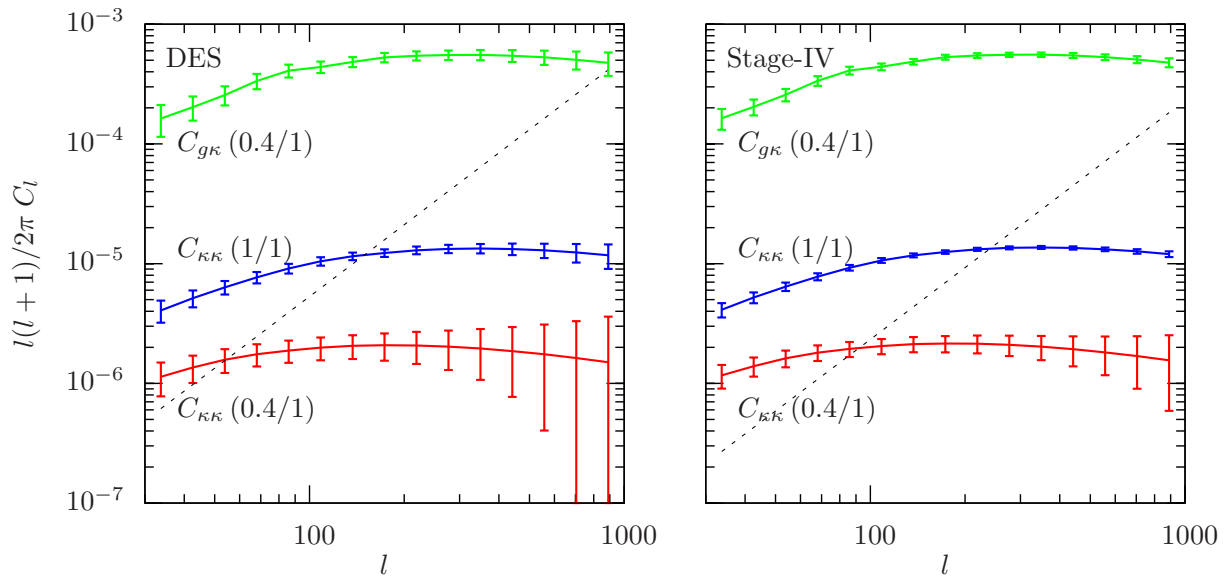}
\caption{
\label{fig:ex_kk_gk} 
Examples of the shear-shear and galaxy-shear power spectra for 
the DES (left panel) and a Stage-IV survey (right panel). 
>From top: the galaxy-shear cross power spectrum $C_{g\kappa}$ with 
foreground galaxies at
$z=0.4$ and background galaxies at $z=1$, the shear-shear auto power 
spectrum  $C_{\kappa \kappa}$. 
The shape noise contribution to the auto power spectrum for $z=1$ is 
shown as well (dashed). 
Note that the shape noise and galaxy shot noise 
contribute to the variance of the power spectra.  
}
\end{figure*}
%

\subsection{Fisher matrix analysis}
\label{errors}

In order to forecast the minimal attainable errors on 
cosmological parameters we implement the Fisher information
approach \citep{vogeley96, tegmark97}
including parameters that describe modifications to gravity.
We treat as observables the following band-power spectra, as 
described in \S \ref{sec:observ}: 
shear-shear power spectra
$\hat{C}_{\kappa_i \kappa_j}(l_m)$, 
galaxy-shear power spectra $\hat{C}_{g_i \kappa_j}(l_m)$
and galaxy-velocity redshift space power spectra $\hat{P}_{gv}(k_m)$. 
The data vector of the projected spectra can be written as 
$\hat{D}_{\nu} = \left\{ \hat{C}_{\kappa_i \kappa_j}(l_m), 
\hat{C}_{g_i \kappa_j}(l_m), \hat{P}_{gv}(z_i, k_n) \right\}$, 
where pairs of tomography bin indices $(i,j):j \geq i$ with 
$i,j=1,...,N_z$ denote independent power spectra. Each of them
is comprised of $N_l$ band-powers. There are also $N_z$ redshift 
space power spectra with $N_k$ band-powers each. 
Therefore, the total  number of observables in the Fisher matrix 
analysis is $N_z(N_z+1) \times N_l + N_z \times N_k$.  
The projected spectra and the redshift space ones are independent so we 
may add their Fisher matrices or merge them into one data vector as above. 

The observable power spectra depend on a set of parameters $p_i$ whose 
uncertainties we aim to forecast. They are the following: 
$\Omega_{m}$ (with $\Omega_{\Lambda}$ 
adjusted to maintain spatial flatness), initial power spectrum slope $n_s$,
normalization of the power spectrum at the epoch of last scattering 
$\Delta^2_{\zeta}(k_0 = 0.002/\mathrm{Mpc})$, and
a bias parameter in each redshift bin $b(z_i)$. 
In addition we use a two parameter ($g_0$, $\eta_0$)  
or one parameter ($\gamma$) 
description of departures from GR 
(see \S \ref{parametrize} for definitions). 
We do not change either the dark matter or baryon physical density when 
other parameters are varied to avoid adding extra information from 
the change of the matter power spectrum shape. 

The Fisher matrix, which measures the curvature of the likelihood 
function in parameter space around its maximum, 
can be expressed for Gaussian distributed observables $\hat{D}_\nu$ as 
\citep{vogeley96}
\begin{equation}
\label{fisher}
	F_{ij} = \sum_{\mu,\nu} \frac{\partial \hat{D}_\mu}{\partial p_i}  
	\mathrm{Cov}^{-1}(\hat{D}_{\mu},\hat{D}_{\nu}) \frac{\partial \hat{D}_\nu}{\partial p_j}. 
\end{equation}
The marginalized $68 \%$-level  error on a parameter $p_i$ is
then given by $\sigma^2(p_i) = \left[ F^{-1} \right]_{ii}$, where $F^{-1}$
is the inverse of the Fisher matrix. 
In order to proceed with forecasting we need to compute the covariance 
matrices in Eqn. (\ref{fisher}). We assume that the band-powers 
constructed from the convergence $\kappa$ and galaxy density $\delta_g$ 
are Gaussian, which restricts the validity of our analysis to relatively
large scales. An analysis of non-Gaussian 
effects on the lensing power spectra may be found in \citep{tj09}. 

The Gaussianity assumption allows us to express the
covariance matrices of the `observed' power spectra as follows
\begin{widetext}
\begin{eqnarray}
\label{eqn:covkk}
\mathrm{Cov} \left[ \hat{C}_{\kappa_i \kappa_j}(l),\hat{C}_{\kappa_m \kappa_n}(l)\right] & = & 
   \left[ \hat{C}_{\kappa_i \kappa_m}(l) \hat{C}_{\kappa_j \kappa_n}(l) 
   + \hat{C}_{\kappa_i \kappa_m}(l) \hat{C}_{\kappa_j \kappa_n}(l) \right]/f_\mathrm{sky} N(l), \\ 
\label{eqn:covgk}
\mathrm{Cov}\left[\hat{C}_{g_i \kappa_j}(l),\hat{C}_{g_m \kappa_n}(l)\right] & = & 
   \left[ \hat{C}_{g_i g_m}(l)  \hat{C}_{\kappa_j \kappa_n}(l) \delta_{im} 
   + \hat{C}_{g_i \kappa_n}(l) \hat{C}_{g_m \kappa_j}(l) \right]/f_\mathrm{sky} N(l), \\
\label{eqn:covkkgk}
\mathrm{Cov}\left[\hat{C}_{g_i \kappa_j}(l),\hat{C}_{\kappa_m \kappa_n}(l)\right] & = & 
   \left[ \hat{C}_{g_i \kappa_m}(l) \hat{C}_{\kappa_j \kappa_n}(l) 
   + \hat{C}_{g_i \kappa_n}(l) \hat{C}_{\kappa_j \kappa_m}(l) \right]/f_\mathrm{sky} N(l), 
\end{eqnarray}
\end{widetext}
where $f_\mathrm{sky}$ is fraction of the sky covered by the survey and 
$N(l) = \sum_{l_{\mathrm{min}}}^{l_{\mathrm{max}}} (2l + 1)$
is the number of independent modes in a passband between $l_{\mathrm{min}}$ and $l_{\mathrm{max}}$.
We assume that  modes in the power spectra are uncorrelated with each other.  

In our analysis the binning in multipoles $l$ is logarithmic -- 
we assume $15$ bins in the range of multipoles 
from $30<l<1000$. We choose this lower limit  in order to be able to apply
the Limber approximation when computing lensing power spectra. The upper limit 
is chosen to limit the nonlinear contributions to the power spectrum. 
Although at the high-$l$ considered there is a nonlinear enhancement to 
the power spectrum, we use the linear contribution as a conservative choice
since both MG effects and biasing can become complex on small scales. 
We will show some results that include the nonlinear enhancement. 

In Fig. \ref{fig:ex_kk_gk} we show example shear-shear and galaxy-shear 
power spectra along with the relevant errors from Eqns. (\ref{eqn:covkk}) 
and (\ref{eqn:covgk}).
The shape noise contribution to the observed spectra is also shown. 
The power spectra are flat as we take only 
the linear evolution of mass perturbations into account. 
The errors on the power spectra scale with $f_\mathrm{sky}^{-1/2}$ 
(see table \ref{surv2}). 

\begin{figure}
\includegraphics[width=16cm]{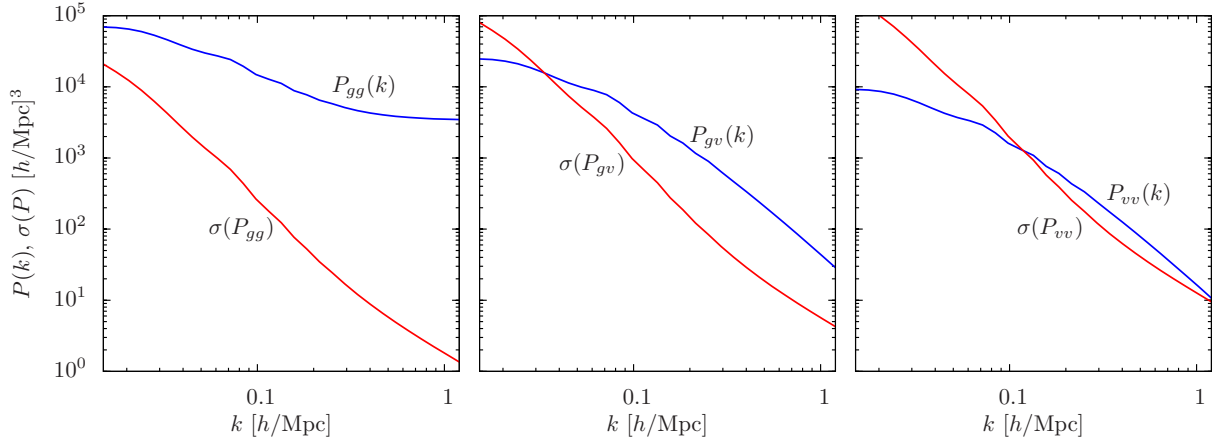}
\caption{
\label{fig:boss_gg_gv_vv} 
The galaxy-galaxy $P_{gg}(k)$ (left panel), galaxy-velocity  $P_{gv}(k)$  (central panel) and velocity-velocity $P_{vv}(k)$
(right panel) power spectra at redshift $0.5<z<0.7$  for the 
BOSS-I survey.
Their expected statistical errors are shown in red. 
We assume $15$ bins per decade in $k$, and do not include nonlinear
effects. 
}
\end{figure}

We are interested in both $\hat{P}_{gv}(k_i)$ and its covariance 
${\rm Cov}[\hat{P}_{gv}(k_i),\hat{P}_{gv}(k_j)]$. 
In order to compute the covariance one could use a standard approach and 
express the covariance by means of the survey's effective volume $V_{\mathrm{eff}}$ as 
$\sigma^2_{\hat{P}_{gv}}(k_i) = 2 V^{-1}_{\mathrm{eff}} \hat{P}^2_{gv}(k_i)$  
\citep{1994ApJ...426...23F, white_song08, song_dore08}. The effective volume  
accounts for the survey window function and in the sample variance limit 
approaches the physical survey volume $V_s$  \citep{white_song08, 
song_dore08}. 
A drawback of this formula is that it underestimates the expected noise of the 
estimator $\hat{P}_{gv}(k_i)$  
and does not account for the effect of discrete sampling 
of redshift space with mass tracers 
(we thank R.~Scoccimarro for drawing our attention to these issues). 
Let us consider the covariance matrix for the galaxy-velocity band-power 
spectrum $\hat{P}_{gv}(k_i)$  defined as
\begin{equation}
{\rm Cov}[\hat{P}_{gv}(k_i),\hat{P}_{gv}(k_j)] = \langle  \hat{P}_{gv}(k_i) \hat{P}_{gv}(k_j) \rangle - P_{gv}(k_i) P_{gv}(k_j). 
\end{equation}
Using the Gaussian assumption this is given by
\begin{equation}
  {\rm Cov}[\hat{P}_{gv}(k_i),\hat{P}_{gv}(k_j)] = 
  \frac{2(2\pi)^3}{2\pi k^2_i \Delta k_i V_s} \delta_{ij} 
  \int^{1}_{-1} \frac{d\mu}{4} \; W_{gv}^2(\mu) P^{(s)}_{gg}(k,\mu).
  \label{eqn:cov1} 
\end{equation}
Next we plug Eqn. (\ref{psz:decomp}) in to Eqn. (\ref{eqn:cov1}) 
and integrate out the angular dependence 
to express the covariance matrix in terms of the component power 
spectra $P_{gg}$, $P_{gv}$, $P_{vv}$ as follows 
\begin{widetext}
\begin{eqnarray}
   {\rm Cov}[\hat{P}_{gv}(k_i),\hat{P}_{gv}(k_j)] 
   & = &
   \frac{2(2\pi)^3}{2\pi k^2_i \Delta k_i V_s}
   \delta_{ij} \frac{105}{18304} \nonumber \\ 
   & \times & \left[ 3003\left(P_{gg}(k_i)+\frac{1}{\bar{n}_g}\right)^2 
   + 4420\left(P_{gg}(k_i)+\frac{1}{\bar{n}_g}\right)P_{gv}(k_i) + 2940P_{gv}^2(k_i)  \right. \nonumber \\
   & + & \left. 10 P_{vv}(k_i)\left\{
   147\left( P_{gg}(k_i)+\frac{1}{\bar{n}_g} \right)+226P_{gv}(k_i)
   \right\}+\frac{24185}{51}P_{vv}^2(k_i)
   \right], \label{pgv:cov1} \\ 
\end{eqnarray}
\end{widetext}
The finite volume of the survey $V_s$ is accounted for as well as the 
shot noise on small scales, which is 
inversely proportional to the mass tracer's density 
$\bar{n}_{\mathrm{g}}$. A 
detailed analysis of the bias and covariance of band-power spectra 
estimators $\hat{P}_{gg}$, $\hat{P}_{gv}$ 
and  $\hat{P}_{vv}$ can be found in the Appendix. 
Formulas for the covariance matrices of these power spectra 
are presented in Eqns. (\ref{pgg:cov}), (\ref{pgv:cov2}) and (\ref{pvv:cov}). 

In Fig. \ref{fig:boss_gg_gv_vv} we show galaxy and velocity  power 
spectra, $P_{gg}(k)$ and $P_{vv}(k)$, 
and the galaxy-velocity cross power spectrum $P_{gv}(k)$
together with their expected errors computed as
described above. The plot shows as an example the power spectra 
from the BOSS-I survey redshift slice centered at $z=0.6$ and $\Delta z=0.2$. 
The errors on $P_{gv}(k)$ are larger by a factor $\sim 4$ from these 
on $P_{gg}(k)$. The good news about the cross power spectrum is that it 
does not suffer from the shot noise which dominates the galaxy power 
spectrum for $k > 0.2 h/\mathrm{Mpc}$. On the other hand it is affected 
by sample variance; for $k <  0.03 h/\mathrm{Mpc}$, $P_{gv}(k)$ has 
limited information. From Fig. \ref{fig:boss_gg_gv_vv} we see that the 
scale dependence of the fractional errors of $P_{gv}(k)$ and $P_{vv}(k)$ 
power spectra bears a typical u-shape with a minimum about 
$k \sim  0.3 h/\mathrm{Mpc}$.
When making error forecasts we limit our calculations to 
$0.015 \; h/\mathrm{Mpc} < k < 0.15 \; h/\mathrm{Mpc}$ to stay within
the linear regime, and use 
$15$ logarithmic band-powers in this wavenumber range. 

We compare uncertainties on redshift-space power spectra to the results of \citep{song_dore08}, whose
work is closely related to ours.
At the fixed scale $k = 0.05 h/\mathrm{Mpc}$ our method yields errors
on $P_{gg}$ which are  $\sim 3$ times smaller than presented in \citep{song_dore08}. 
On the other hand errors on $P_{vv}$ in our analysis are $\sim 2.5$ larger than those in \citep{song_dore08}.

The total signal-to-noise ratio (S/N) for $P_{gv}(k)$ is presented in 
Fig. \ref{fig:s2n} as a function of redshift. 
We also show the 
$S/N$ for the galaxy-shear power spectra $C_{g \kappa}$ for 
the DES and a Stage-IV surveys assuming that 
source galaxies are at fixed redshift -- $z_s=1.1$ for DES and $z_s=1.9$ 
for the Stage-IV. 
As expected, the S/N is the highest when lensing galaxies are about 
half way to the source galaxies. The most robust constraints on MG 
are expected when an imaging and spectroscopic survey both have high 
S/N over a common redshift range. For DES and BOSS-I this occurs at 
redshifts between about 0.4-0.5. 

It is worth noting that the S/N for $P_{gv}(k)$  is inversely 
related to the bias $b$ of the galaxy sample. 
A joint analysis by 
combining different population galaxy samples may be helpful 
in beating down the sample variance on large scales \cite{mcdonald09}.

\subsection{Surveys}
\label{surveys}

\begin{figure}
\includegraphics[width=14cm]{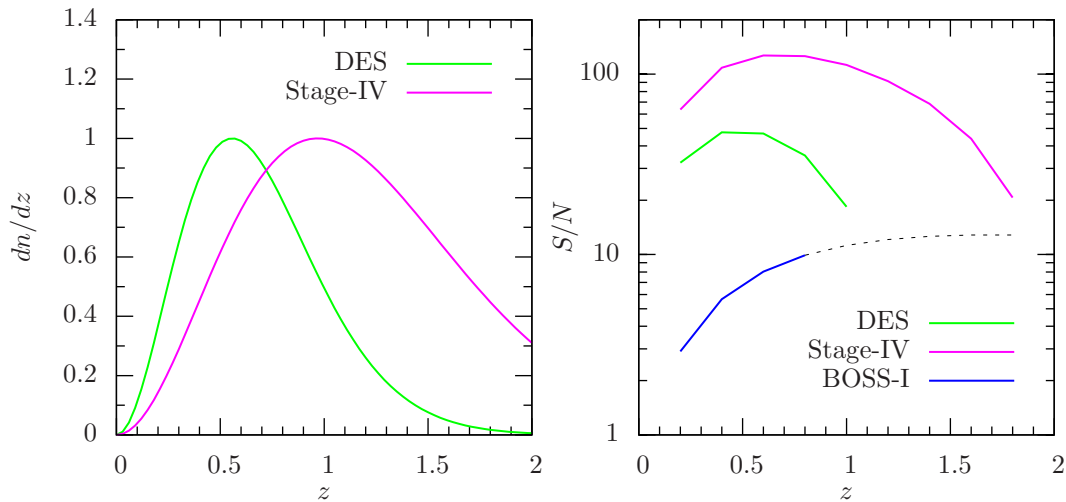}
\caption{
\label{fig:s2n} 
Left panel:
Redshift distribution of galaxies for imaging surveys 
specified in Table \ref{surv2}. The distributions are normalized
so that the maximum value is unity.   
Right panel: 
The total signal-to-noise ratio for the galaxy-shear power spectra $C_{g \kappa}$ as a function of
redshift of lensing galaxies. The background galaxies are located in
the furthest bin --  $z=1.2$ for DES,  
$z=2.0$ for a Stage-IV survey. The galaxy-velocity power spectrum
signal-to-noise  
is also shown for a BOSS-I-type survey (extended to
$z=1.9$ with the same sky coverage and   galaxy density as the dotted curve).  
}
\end{figure}

One of the main scientific goals of upcoming multi-color 
imaging surveys is to measure cosmological weak gravitational lensing.
We consider two surveys of this kind: 
the Dark Energy Survey (DES) \citep{des} which is expected to begin data 
acquisition in 2011, and a generic Stage-IV survey \citep{detf} whose 
example is the LSST survey \citep{lsst}. 

The surveys are characterized by 
sky coverage $f_{\mathrm{sky}}$, 
surface density of lensed galaxies $n^{\mathrm{2d}}_{\mathrm{g}}$ and 
the galaxy redshift distribution. The sky coverage for
the DES is taken to be 
$5000$ sq. degs. and for the Stage-IV survey $20000$ sq. degs.  
The redshift distribution of galaxies in the imaging surveys is 
assumed to have the form 
$dn/dz \propto z^2 \: \mathrm{exp}\left[-\left(z/z_0\right)^{3/2} \right]$, 
where the values of $z_0$ for surveys under consideration are given 
in the table \ref{surv2}  and the 
distributions are shown in the Fig. \ref{fig:s2n}. 

In order to measure the redshift-space power spectrum $P^{(s)}_{gg}$  
we consider spectroscopic surveys. The Baryon Oscillation
Spectroscopic Survey (BOSS) \citep{boss} will target Luminous Red 
Galaxies (LRG) up to redshift $z \sim 0.7$ and will cover a quarter of the sky.
It will obtain spectra of $1.5\times10^6$ galaxies, which 
results in the number density  
$\bar{n}_{\mathrm{g}}= 1.1\times10^{-4} \mathrm{Mpc}^{-3}$. 
Based on the LRG sample from the SDSS survey one expects these objects
to be biased by a factor of $b \simeq 2$ with respect to the mass 
distribution. We assume that galaxies 
are uniformly distributed across the redshift range. 
In addition to the BOSS survey (called BOSS-I throughout the paper) 
we consider a futuristic version 
(dubbed BOSS-II here) with double the sky coverage compared to 
BOSS-I (to keep up with the sky coverage of Stage-IV survey), 
the same galaxy number density, and extending to 
redshift $z=1.1$. 

\begin{table}
\begin{tabular}[b]{r  c c c c c c }
\hline
	& $f_{\mathrm{sky}}$ & $n^{\mathrm{2d}}_{\mathrm{g}}$ & $z_0$ & $\left< z \right>$ \\
\hline 
DES      &  $ 5000 $  	& $15$ & $0.46$ & $0.7$ \\ 
Stage-IV & $ 20000 $	& $30$ & $0.8$  & $1.2$ \\ 	
\hline
\end{tabular}
\caption{
\label{surv2}
Parameters of imaging surveys: 
sky coverage $f_{\mathrm{sky}}$ in sq. degs., galaxy surface 
density $n^{\mathrm{2d}}_{\mathrm{g}}$ per sq. arcmin., 
the $z_0$ parameter of the galaxy redshift distribution, and  
its mean $\left< z \right>$. 
}
\end{table}

\begin{table}
\begin{tabular}[b]{r  c c c c c c }
\hline
	& $f_{\mathrm{sky}}$ & $z_{\mathrm{min}}$ & $z_{\mathrm{max}}$ & $V_s$ & $\bar{n}_{\mathrm{g}}$ & $n^{\mathrm{2d}}_{\mathrm{g}}$ \\
\hline 
BOSS-I 		&  $ 10000 $  	& $0.1$ & $0.7$  & $15.5$ & $1.1 \times 10^{-4}$ & $0.05$ \\ 
BOSS-II		&  $ 20000 $	& $0.1$ & $1.1$  & $90$  & $1.1 \times 10^{-4}$ & $0.14$ \\ 	
\hline
\end{tabular}
\caption{
\label{surv3}
Parameters of the spectroscopic surveys: sky coverage $f_{\mathrm{sky}}$ in sq. degs., minimum and maximum redshift limits of 
the survey, $z_{\mathrm{min}}$ and $z_{\mathrm{max}}$,
survey comoving volume in Gpc$^3$, mean spatial density of galaxies  $\bar{n}_{\mathrm{g}}$ per Mpc$^3$ and their mean projected density 
$n^{\mathrm{2d}}_{\mathrm{g}}$ per sq. arcmin.
}
\end{table}

\section{Results}
\label{sec:results}

We quantify constraints on modifications to
gravity by combined measurements from upcoming imaging 
and spectroscopic surveys. 
Imaging surveys (see Table \ref{surv2}) provide us with the lensing 
signal via $C_{\kappa \kappa}(l)$ and 
and $C_{g \kappa}(l)$. The lensing power spectra  
depend on the modified gravity parameters and ($C_{g \kappa}(l)$ only) 
on redshift dependent galaxy bias, as given by Eqn. (\ref{cls2}). 
The redshift space power spectrum $P_{gv}(k)$ can be obtained from 
spectroscopic surveys. It depends on the ratio of $g_0$ and $\eta_0$ 
and on galaxy bias.

Our fiducial cosmological model is given by 
$\Omega_{m}=0.26$, the present
value of the Hubble constant $H_0 = h \times 100 \mathrm{km/s/Mpc} = 
72 \mathrm{km/s/Mpc}$, logarithmic slope of the initial 
matter power spectrum $n=0.96$ and its amplitude 
$\Delta^2_{\zeta}(k_0 = 0.002/\mathrm{Mpc}) = 2.41 
\times 10^{-9}$ \citep{wmap5}.
The redshift dependent galaxy bias is a free parameter in each 
redshift bin with a fiducial value $b=2$, corresponding to
the Luminous Red Galaxy (LRG) sample from the SDSS. 
Fiducial values for parameters which describe  modifications to 
gravity are set to their values in GR: $g_0=\eta_0=1$ and $\gamma=0.55$. 
We use statistical priors on $\Omega_m$, $\Omega_m h^2$,  $\Omega_b h^2$, 
and the power spectrum  parameters  
$\Delta^2_{\zeta}$ and $n$ as expected from the 
Cosmic Microwave Background measurements by the 
\emph{Planck} satellite \citep{Takada2004}.
We also assume that the 
distance-redshift relation is unchanged
from the standard $\Lambda$CDM cosmology. Therefore, MG enters the 
equations through the growth of structure
and influences the weak lensing and redshift space power spectra, 
but does not affect the distance-redshift relation. 
Last but not least, we present uncertainties on cosmological parameters 
after uncertainties in the other parameters were marginalized out. 

In our analysis we use tomographic measurements which require binning of 
the lensing and redshift-space power spectra 
in redshift intervals. We assume bins with width $\Delta z = 0.2$. 
Thus for the lensing spectra
from the DES survey we use  $6$ redshift bins, while for 
for the  Stage-IV survey we use $10$ bins. For spectroscopic surveys we 
use $3$ redshift bins for BOSS-I  and $5$ for BOSS-II. 
The projected density of spectroscopic galaxies is smaller by a factor 
of a few tens than for the imaging surveys, 
as given in tables \ref{surv2} and \ref{surv3}. 
This affects the errors on $C_{gg}$ large but does not effect 
$C_{g \kappa}$ which is one of our observables. 

\begin{figure*}
\includegraphics[width=14cm]{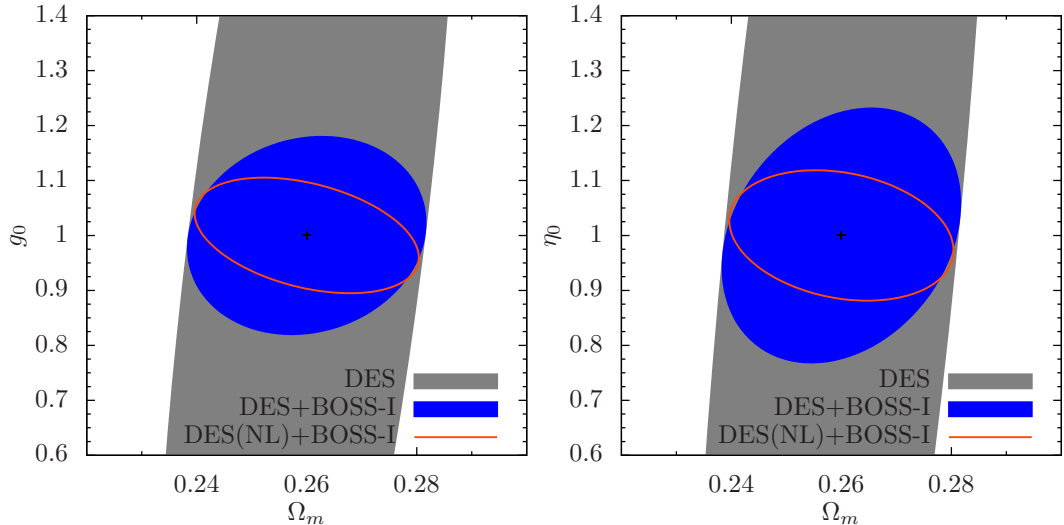}
\caption{
\label{fig:des_boss_geta} Forecast constraints on 
the effective gravitational constant $g_0$, 
the ratio of metric potentials $\eta_0$, and 
$\Omega_m$ for the DES imaging survey and the BOSS-I redshift survey.
Forecasts based on the linear lensing power spectra 
(shear-shear and galaxy-shear) for DES are shown 
in light gray, and combined with the BOSS-I galaxy-velocity 
power spectrum in dark gray (blue). 
Light (red) inner contours show the forecasts if the 
nonlinear lensing power spectrum is used. 
Other parameters are marginalized over, and all contours 
show the $68\%$ confidence level. 
}
\end{figure*}

Our results on MG parameters constraints for the DES and BOSS-I surveys
are shown in Fig. \ref{fig:des_boss_geta}. 
If we consider weak lensing observables only (with CMB priors
as described above) the one-sigma error on $g_0$ and $\eta_0$ is 
$\sigma(g_0)=0.94$ and $\sigma(\eta_0)=1.2$. The weak constraints are due to the
strong covariance between these parameters and redshift dependent bias. 
However, when combined with BOSS-I the constrains on $g_0$ and $\eta_0$ 
improve to $\sigma(g_0)=0.12$ and $\sigma(\eta_0)=0.15$, as 
shown in the Fig. \ref{fig:des_boss_geta}. Thus redshift-space 
clustering data enables us to beat down errors on MG parameters by about
a factor of 8 (note that the constraints on 
$\Omega_{m}$ come mainly from the CMB prior). 
The BOSS-I survey as presently planned will obtain redshifts for 
objects with $z<0.7$.  
With the expanded redshift survey BOSS-II, 
the accuracy would improve to $\sigma(g_0)=0.086$ and $\sigma(\eta_0)=0.11$. 

In order to use information on small scales from lensing 
one needs to model the nonlinear evolution of the matter density. 
Nonlinear evolution boosts the lensing signal $C_{\kappa \kappa}(l)$ 
by a factor of $\sim 4$ on scales $l \sim 500-1000$.  
Numerical simulations which 
provide nonlinear matter power spectra in MG theories 
are in their infancy but the 
first attempts are encouraging, as shown by \citep{LimaHu} for 
the $f(R)$ models and \citep{khoury09,schmidt09} for the DGP model.
However, there is no simple and model-independent parametrization of 
the nonlinear corrections to the matter power spectrum in MG models. 
This is due to the complexity of the evolution equations and the existence
of additional fields that drive the theories to GR on small scales.  

Therefore we use only the linear power spectrum 
up to $l=1000$. This underestimates the signal-to-noise in lensing. 
An alternative is to use the GR-based nonlinear power spectrum,  
as our fiducial model is GR and small deviations from the GR may not 
introduce substantial deviations in the nonlinear evolution.
In Fig. \ref{fig:des_boss_geta} we also show predicted uncertainties 
in the MG parameters if we include such a nonlinear power spectrum 
in the modeling of the lensing spectra. The errors 
on both $g_0$ and $\eta_0$ drop by almost a factor of $2$ 
to $\sigma(g_0)=0.069$ and $\sigma(\eta_0)=0.078$, respectively, 
compared to the linear case. 
This improvement is partly due to breaking the degeneracy between
redshift dependent bias and MG parameters in the nonlinear regime. 

\begin{figure*}
\includegraphics[width=14cm]{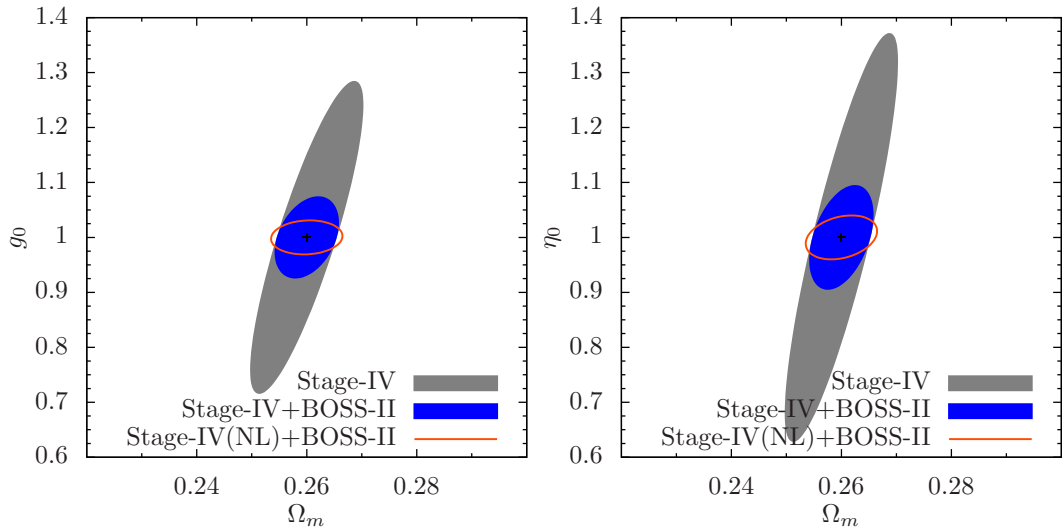}
\caption{
\label{fig:stage4_boss2_geta} 
Forecasts as in Fig. \ref{fig:des_boss_geta} but for a Stage-IV
imaging survey and the BOSS-II spectroscopic survey. 
}
\end{figure*}

We also examine Stage-IV-type surveys like LSST, which will 
have greater depth and sky coverage. 
The predictions are presented in  Fig. \ref{fig:stage4_boss2_geta}.  
Without spectroscopic information the constraints are 
$\sigma(g_0)=0.19$ and $\sigma(\eta_0)=0.24$. 
If we combine imaging data with the redshift-space 
power spectrum $P_{gv}(k)$ from the BOSS-II survey, 
the constraints improve to $\sigma(g_0)=0.048$  and 
$\sigma(\eta_0)=0.062$. Including nonlinear evolution
in the lensing spectra leads to about a factor of two improvement. 

\begin{figure}
\includegraphics[width=14cm]{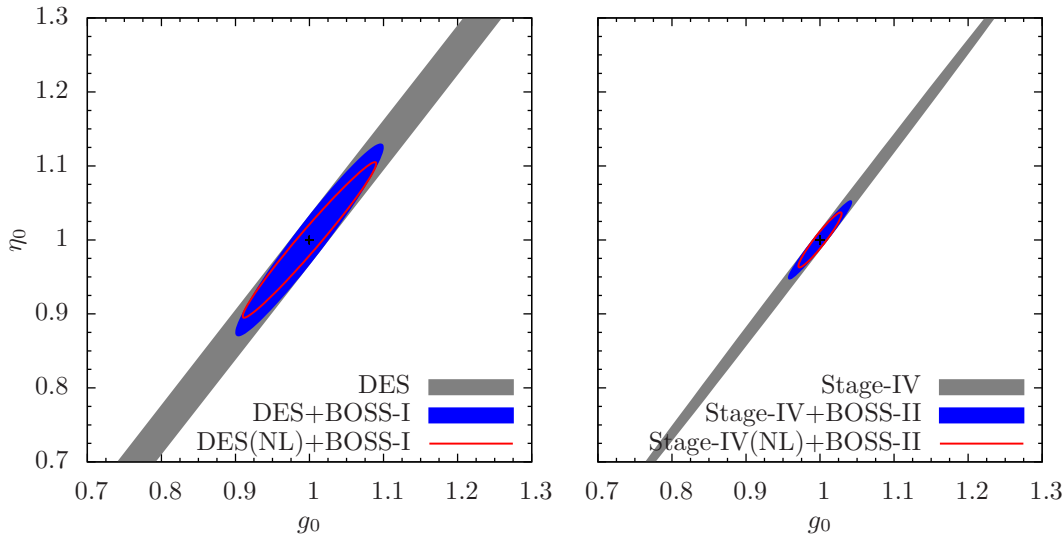}
\caption{
\label{fig:gmu_corr} 
Correlation between modified gravity parameters for the DES and 
BOSS-I surveys (left panel), and for Stage-IV surveys (right panel).
The error contours are as in Fig. \ref{fig:des_boss_geta}. 
}
\end{figure}

The MG parameters are strongly correlated with each other as shown 
in Fig. \ref{fig:gmu_corr}. This is mostly due to the 
dependence of the growth factor solely on the ratio of the MG parameters. 
The dependence of the lensing power spectra on a different 
combination of MG parameters (see Eqn. (\ref{cls2})) 
is not a strong effect in practice; 
moreover, the change in growth can be compensated by a change in 
the redshift dependent galaxy bias which substantially enhances 
correlations. For a different MG parametrization like the one 
presented in the 
Appendix \ref{app:mg} one expects a weaker degeneracy. 
By combining lensing data with the redshift-space 
 $P_{gv}(k)$ the MG parameters 
are more reliably determined, as their correlation with the galaxy bias 
parameters is significantly reduced (see 
Eqns. (\ref{cls2}) and  (\ref{eqn:pgv})).

The importance of breaking the 
degeneracy between the galaxy bias and the growth of structure 
is highlighted in Fig.~\ref{fig:gmu_corr} by considering the 
dependence of uncertainties in $g_0$ and $\eta_0$ on the 
limiting projected scale $l_\mathrm{max}$ in the lensing power 
spectra  and the limiting 
physical scale $k_{\mathrm{max}}$ in the $P_{gv}(k)$ power spectrum. 
For DES and BOSS-I we find that 
these constraints are insensitive 
to $l_\mathrm{max}$ if linear spectrum is used;  
the correlation coefficient between $g_0$ and galaxy bias is  
$r(g_0,b_i) \sim 0.8$.  
The degeneracy can be broken by 
using the nonlinear lensing power spectra which introduces scale 
dependent growth,
improving the constraints by a factor of $\sim 2$ and 
lowering the correlation coefficient to $r(g_0,b_i) \sim 0.5$.
Another way to beat down errors is to increase the 
range of $k$-modes in the redshift space power spectrum.
However, extending the range of $k$-modes to $k_\mathrm{max} > 0.15$
requires modeling 
nonlinear evolution and velocity dispersion effects. 
Throughout the paper we use linear lensing  power spectra  with $l_\mathrm{max}=1000$ and linear 
redshift space power spectrum with $k_\mathrm{max}=0.15$~h/Mpc. 
In this case $r(g_0,b_i) \sim 0.8$.

The bias value of a given galaxy sample is related to the S/N ratio for 
the redshift space power spectrum,   
as discussed in \S \ref{errors}. Lower bias implies higher 
signal-to-noise for $P_{gv}(k)$,
which is counterbalanced by a lower S/N for 
$C_{g \kappa}$ in the joint analysis. 
The net effect is that for the fiducial bias $b=1$ the constraints on 
$g_0$ and $\eta_0$ are virtually unchanged
compared to the $b=2$ case which we have used in the analysis. 
If a less biased sample with $b=0.5$ is chosen,  
the predicted errors increase by $3\%$. 
The fiducial bias value is therefore unimportant for the purpose of this work. 

\begin{figure*}
\includegraphics[width=14cm]{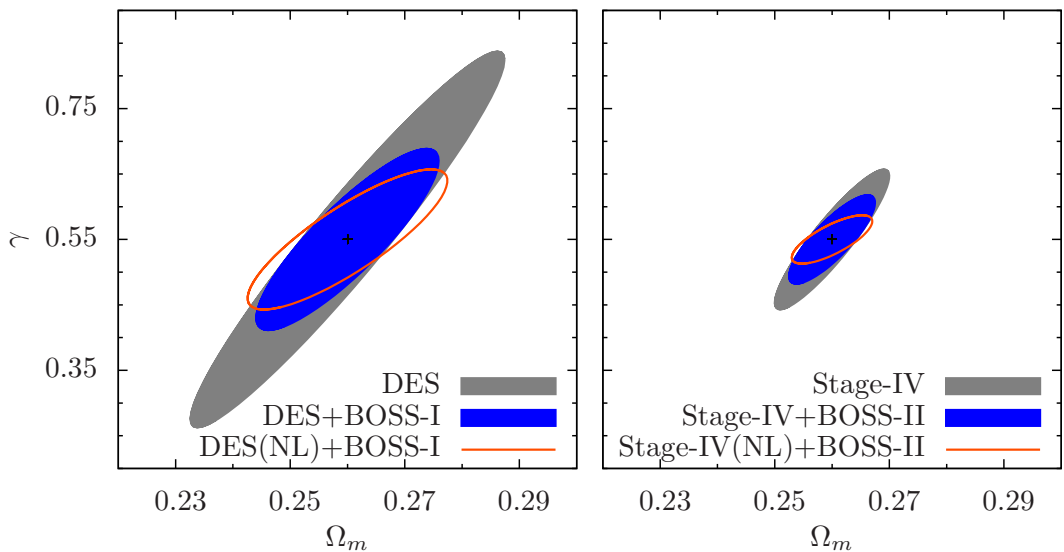}
\caption{
\label{gamma_all} Achievable errors in the growth parameter $\gamma$ around a fiducial value $\gamma=0.55$.
There are shown  $68\%$ confidence level contours for survey configurations as described in  
Figs. \ref{fig:des_boss_geta} and \ref{fig:stage4_boss2_geta}. 
There are shown $68\%$ confidence level contours as expected from the DES and Stage-IV imaging surveys (gray) 
and when combined with spectroscopic surveys (blue). The contour (orange) shows predictions when the nonlinear matter power
spectrum is accounted for. 
The braneworld DGP model has $\gamma=0.68$.  
}
\end{figure*}

We have used the $(g_0,\eta_0)$ parametrization of MG so far. 
Another parametrization, based on the growth exponent $(\gamma)$,  
has been shown to be useful in distinguishing between 
gravity models \citep{huterer_linder07}. 
Our results for $\gamma$ are presented  in Fig. \ref{gamma_all}.
It shows that using DES lensing data we expect to constrain $\gamma$ 
to $\sigma(\gamma)=0.19$,  
while Stage-IV survey lensing data achieve $\sigma(\gamma)=0.07$.  
When information about the redshift space power spectrum  
$P_{gv}(k)$ is included the improvement is about a factor of 
two for DES combined with BOSS-I, and less than that for Stage IV surveys. 
Thus combining imaging and spectroscopic surveys is less 
useful for constraining $\gamma$ than the parametrization used
in the rest of the paper (compare Fig.~7 with Figs.~5 and 6). 
This is not surprising, as 
any measure of the growth of structure constrains this one-parameter 
modification. While it may indeed capture the relevant physics in some 
MG models, in general robust constraints require tests of both 
the Poisson equation and the ratio of metric potentials. These require 
both lensing and dynamical information, as illustrated in Figs.~5 and 6. 

The two models that have been extensively worked out in the literature 
are $f(R)$ and DGP models; for these the ratio of potentials departs
from unity by tens of percent in the quasi-static, Newtonian regime
relevant to large-scale structure. The
gravitational constant however is close to its value in GR if it is
defined using the sum of metric potentials in the Poisson equation 
(rather than the usual definition, used in this paper as well, with
the curvature potential) \citep{HuSaw}. Thus $\eta_0$ is the more sensitive
parameter for testing MG if current models are used as a guide. 

\section{Conclusions}
\label{sec:conclusions}

We have used a Fisher matrix approach to test 
modified gravity models using imaging and spectroscopic galaxy surveys.
The expansion history universe is not likely to be sufficient
to test MG models as it could be mimicked by an appropriately evolving
dark energy equation of state.  
Here we use observable consequences of the evolution of perturbations
to test gravity. In particular we use 
weak gravitational lensing measured from multi-color imaging surveys
and dynamical information from spectroscopic surveys of galaxies. 
Lensing is sensitive to both metric potentials, whereas dynamical 
effects are driven by the Newtonian potential. Combining these probes 
provides robust tests of gravity, which is valuable given that we 
currently have a very limited number of specific models that are at 
all plausible. 

We use three simple parametrization of  MG models and 
perform a joint analysis of shear-shear, 
galaxy-shear and galaxy-velocity power spectra. 
With a two-parameter description of modified gravity, we find
that combining the three observables is essential to obtain strong constraints 
on gravity. We give predictions for Stage III and Stage IV surveys. 
We also compare the two-parameter description to the commonly used 
single parameter description (via the $\gamma$ parameter), and to 
a scale dependent description used for braneworld models. 

Our results highlight the need for imaging and spectroscopic surveys
to probe the same redshift range (for the lensing mass and galaxy 
distribution respectively). Planned imaging surveys that reach 
redshifts of unity and beyond, and spectroscopic surveys that measure
galaxy clustering at $z\sim 0.2-0.6$ are already well suited for 
probing modified gravity. With a careful selection of galaxy samples 
to compare cross-spectra, such surveys will allow us to perform robust tests of
gravity. The question of how to use small scale information requires 
significant work, as different modified gravity models show a variety of 
nonlinear effects on scales below $\sim 10$ Mpc. It may be that 
models will need to be tested individually on these scales. 
Even so, measurements of the two parameters we have used would provide a 
consistency test of the GR plus a smooth dark energy scenario 
over a wide range of scales. 

We have focused on the most common MG parametrization with the 
gravitational constant and the ratio of the metric potentials being 
free parameters. These parameters
turn out to be strongly correlated with each other and with 
the evolving, scale independent bias for the 
observables we have used. With external information on bias, or by including
galaxy-galaxy spectra, this degeneracy could be broken (though 
the bias parametrization would need to be more complex as well with
the inclusion of scale dependence).  
Alternatively, a different, physically-motivated parametrization 
may be better able to capture the dependence of observables 
on MG, such as the one discussed in the Appendix \ref{app:mg}
\citep{amin, zhao08}.
The general properties of a useful  MG parametrization, 
one that is able to get 
the most information out of a given set of observables, 
have been studied by \citep{zhao09}. 

Finally, in our analysis when we have included nonlinear evolution
we have simply assumed it follows the predictions for GR. 
This provides one scenario for MG constraints; it may be optimistic
as the degeneracy between MG parameters and scale independent galaxy 
bias gets lifted in this case. The existence of scale dependent 
bias on scales where nonlinear effects are important would lead to 
weaker constraints. On the other hand specific signatures of non-linearity
would make it easier to distinguish models. 
Clearly more work is needed to include small scale information, 
realistic biasing
schemes and additional observables such as galaxy power spectra, 
CMB lensing and the ISW effect. 


\begin{acknowledgments}
We are grateful to 
Gary Bernstein, Alex Borisov, Mike Jarvis, Marcos Lima and 
Pengjie Zhang for many useful discussions. We especially thank
Roman Scoccimarro for discussions and help with the covariance
calculations. This work is
supported in part by NSF grant AST-0607667.
MT is supported by World Premier International Research Center
Initiative (WPI Initiative), MEXT, Japan, by Grand-in-Aid for Scientific
Research on Priority Area No. 467 ``Probing Dark Energy through an
Extremely Wide and Deep Survey with Subaru Telescope'' and on young
researchers (Nos. 17740129 and 20740119). 
\end{acknowledgments}


\appendix
\section{Covariance matrices for the redshift-space  power spectra}
\label{app:name}

We consider distribution of mass tracers (galaxies) in the redshift space $\delta^{(s)}_g$ and its power spectrum 
$P^{(s)}_{gg}$ which is defined as \citep{2002PhR...367....1B}
\begin{equation} 
   \langle \delta^{(s)}_g(\bm{k}) \delta^{(s)}_g(-\bm{k'}) \rangle = 
   (2 \pi)^3  P^{(s)}_{gg}(\bm{k}) \delta_D(\bm{k} - \bm{k'}),
   \label{app:psdef}
\end{equation} 
where $\delta_D(\bm{k})$ is the Dirac's delta function. 
In the discrete case (\ref{app:psdef}) becomes 
\begin{equation} 
   \langle \delta^{(s)}_g(\bm{k}_i) \delta^{(s)}_g(-\bm{k}_j) \rangle = 
   \frac{(2 \pi)^3}{V_F}  P^{(s)}_{gg}(\bm{k}_i) \delta_{ij},
   \label{app:psdisc}
\end{equation} 
where discrete Fourier modes $\delta^{(s)}_g(\bm{k}_i)$ have units of volume. 
We are interested in computing bias and covariance matrices for the 
galaxy-galaxy band-power spectrum $P_{gg}(k_i)$, galaxy-velocity $P_{gv}(k_i)$ and velocity-velocity $P_{vv}(k_i)$.
First, let us define estimators of these power spectra 
\begin{equation} 
	\hat{P}_{XY}(k_i) =  \frac{1}{N_k} \sum_{k,\mu} W_{XY}(\mu) \hat{P}^{(s)}_{gg}(k,\mu), \label{app:psestim}
\end{equation} 
where $X$ and $Y$ stand for galaxy ($g$) or velocity ($v$) fields. The weight functions $W_{XX}(\mu)$ 
are given by
\begin{eqnarray}
	W_{gg}(\mu) &=& P_0(\mu) - \frac{5}{2}P_2(\mu)+\frac{27}{8}P_4(\mu),  \label{app:wgg} \\
	W_{gv}(\mu) &=& \frac{15}{4}P_2(\mu)-\frac{135}{8}P_4(\mu),  \label{app:wgv} \\ 
	W_{vv}(\mu) &=& \frac{315}{8}P_4(\mu), \label{app:wvv}
\end{eqnarray}
where $P_l(\mu)$ is the $l$-th order Legendre polynomial as a function of the azimuthal angle between a wavevector
$\bm{k}$ and a line of sight ($\mu_k \equiv k_{\parallel}/k$). The expression (\ref{app:wvv}) 
agrees with the one found by \citep{2008PhRvD..78j3512S}. By ensemble averaging of 
(\ref{app:psestim}) one can show that weight functions (\ref{app:wgg}), (\ref{app:wgv}) and 
(\ref{app:wvv})  together with (\ref{app:psestim})
provide unbiased estimators for $P_{gg}(k_i)$, $P_{gv}(k_i)$ and $P_{vv}(k_i)$. 
In the continuous limit, if the fundamental volume $V_F = (2\pi)^3/V_s$ is small compared to the total 
volume of the Fourier-space spherical shell $V_k$ where averaging is carried on, we obtain 
\begin{equation}
	\langle \hat{P}_{XY}(k_i) \rangle  = \frac{1}{N_k} \frac{1}{V_F} 
        \int dk \; 2\pi k^2 \int^{1}_{-1} d\mu \; W_{XY}(\mu) \left[ P_{gg}(k_i) + 2 \mu^2 P_{gv}(k_i) + \mu^4 P_{vv}(k_i)\right].  
\end{equation}
Next, we use orthogonality relation for Legendre polynomials and after integration obtain 
\begin{equation}
	\langle \hat{P}_{XY}(k_i) \rangle  = \frac{1}{N_k} \frac{V_k}{V_F} P_{XY}(k_i) = P_{XY}(k_i).
\end{equation}
Therefore estimators (\ref{app:psestim}) of power spectra $P_{gg}$, $P_{gv}$ and $P_{vv}$ are unbiased.  

Now let us turn to computing covariance matrices for the introduced estimators. The formulation
is general for all three band-power spectra we are interested in. The difference is 
in the weight functions. 
>From the  definition of the covariance matrix we have
\begin{equation}
   {\rm Cov}[\hat{P}_{XY}(k_i),\hat{P}_{XY}(k_j)] = \langle  \hat{P}_{XY}(k_i) \hat{P}_{XY}(k_j) \rangle 
   - P_{XY}(k_i) P_{XY}(k_j). 
   \label{app:cov_def}
\end{equation} 
We plug Eqn. (\ref{app:psestim})  in Eqn. (\ref{app:cov_def}) and obtain 
\begin{eqnarray}
   {\rm Cov}[\hat{P}_{XY}(k_i),\hat{P}_{XY}(k_j)] & + & P_{XY}(k_i) P_{XY}(k_j)  \nonumber \\
   & = & \frac{1}{N_k^2} \sum_{k,\mu} \sum_{k',\mu'} W_{XY}(\mu) W_{XY}(\mu') 
   \langle \hat{P}^{(s)}_{gg}(k,\mu) \hat{P}^{(s)}_{gg}(k',\mu'), \rangle 
\end{eqnarray}
where averaging is over all modes is spherical shells of radii $k$ and $k'$ in the Fourier space. 
We can express the estimator of the redshift space power spectrum by means of (\ref{app:psdisc})
which leads to 
\begin{eqnarray}
   {\rm Cov}[\hat{P}_{XY}(k_i),\hat{P}_{XY}(k_j)]  & + & P_{XY}(k_i) P_{XY}(k_j)  \nonumber \\
   & = & \frac{1}{N_k^2} \frac{V_F^2}{(2\pi)^6} \sum_{k,\mu} \sum_{k',\mu'} W_{XY}(\mu) W_{XY}(\mu') 
   \langle \delta^{(s)}_g(\bm{k}) \delta^{(s)}_g(-\bm{k}) \delta^{(s)}_g(\bm{k'}) \delta^{(s)}_g(-\bm{k'}) \rangle .   
   \label{app:psdd}
\end{eqnarray}
Note that $\bm{k}$ denotes discrete Fourier modes contained in a  spherical-shell region of the Fourier space 
for a given band-power $k_i$:  $k_i-\Delta k_i/2\le |\bm{k}|\le k_i+\Delta k_i/2$. 
We assume that the $\delta^{(s)}_g(\bm{k})$ is the Gaussian random field
which allows to simplify (\ref{app:psdd}) considerably by applying the Wick's theorem (see e.g. 
\citep{2002PhR...367....1B}). We obtain 
\begin{eqnarray}
   {\rm Cov}[\hat{P}_{XY}(k_i),\hat{P}_{XY}(k_j)] &+&  P_{XY}(k_i) P_{XY}(k_j)   \nonumber  \\
   & = & \frac{1}{N_k^2} \frac{V_F^2}{(2\pi)^6} \sum_{k,\mu} \sum_{k',\mu'} W_{XY}(\mu) W_{XY}(\mu') 
   \left[ \langle \delta^{(s)}_g(\bm{k}) \delta^{(s)}_g(-\bm{k}) \rangle  
   \langle \delta^{(s)}_g(\bm{k'}) \delta^{(s)}_g(-\bm{k'}) \rangle \right.  \nonumber \\
   & + &  \left. \langle \delta^{(s)}_g(\bm{k})  \delta^{(s)}_g(\bm{k'}) \rangle 
   \langle \delta^{(s)}_g(-\bm{k}) \delta^{(s)}_g(-\bm{k'}) \rangle +
   \langle \delta^{(s)}_g(\bm{k})  \delta^{(s)}_g(-\bm{k'}) \rangle 
   \langle \delta^{(s)}_g(-\bm{k}) \delta^{(s)}_g(\bm{k'}) \rangle \right] .
\end{eqnarray}
By applying relation (\ref{app:psdisc}) we obtain 
\begin{eqnarray}
   {\rm Cov}[\hat{P}_{XY}(k_i),\hat{P}_{XY}(k_j)] &+&  P_{XY}(k_i) P_{XY}(k_j) \nonumber \\ 
   &=& \frac{1}{N_k^2} \sum_{k,\mu}  W_{XY}(\mu) P^{(s)}_{gg}(k,\mu) \sum_{k',\mu'} W_{XY}(\mu') P^{(s)}_{gg}(k',\mu') \nonumber \\
   &+& \frac{1}{N_k^2} \sum_{k,\mu} \sum_{k',\mu'}  W_{XY}(\mu) W_{XY}(\mu') [P^{(s)}_{gg}(k,\mu)]^2 \delta_{k,k'} \delta_{\mu,-\mu'} \nonumber \\
   &+& \frac{1}{N_k^2} \sum_{k,\mu} \sum_{k',\mu'}  W_{XY}(\mu) W_{XY}(\mu') [P^{(s)}_{gg}(k,\mu)]^2 \delta_{k,k'} \delta_{\mu,\mu'}. 
\end{eqnarray}
In the second and third terms we introduced Kronecker delta-type symbol $\delta_{p,q}$ which means that 
only pairs of modes which wavevectors are opposite contribute to the second term and only these
which wavevectors are equal contribute to the third term. 
These delta functions make one summation in the second and third terms drop out. Moreover, all functions
are even with respect to $\mu$. Next, in the limit of continuous $\mu$  we derive 
\begin{equation}
   {\rm Cov}[\hat{P}_{XY}(k_i),\hat{P}_{XY}(k_j)] = 
   \frac{2}{V_F} \frac{1}{N_k^2} \delta_{ij} \sum_{k} 2\pi k^2 \Delta k 
  \int^{1}_{-1} d\mu \; W_{XY}^2(\mu) [P^{(s)}_{gg}(k,\mu)]^2.
\end{equation}
Finally, we obtain a general expression for the covariance of the $P_{gg}(k_i)$, $P_{gv}(k_i)$ or $P_{vv}(k_i)$
provided validity of the decomposition (\ref{psz:decomp}) 
\begin{equation}
   {\rm Cov}[\hat{P}_{XY}(k_i),\hat{P}_{XY}(k_j)] = 
   \frac{2(2\pi)^3}{2\pi k^2_i \Delta k_i V_s} \delta_{ij} 
   \int^{1}_{-1} \frac{d\mu}{4} \; W_{XY}^2(\mu) P^{(s)}_{gg}(k,\mu). 
   \label{app:main}
\end{equation}
By means of the relation (\ref{app:main}) it is straightforward to obtain
desired expressions for the $\hat{P}_{gg}(k_i)$,  $\hat{P}_{gv}(k_i)$ and $\hat{P}_{vv}(k_i)$ 
band-power-spectra covariance matrices which are presented below. 
Notice that we included the effect of discrete sampling of the mass tracer distribution
by including the shot noise term $1/\bar{n}_g$. 
\begin{widetext}
\begin{eqnarray}
   {\rm Cov}[\hat{P}_{gg}(k_i),\hat{P}_{gg}(k_j)]
   & = & 
   \frac{2(2\pi)^3}{2\pi k^2_i \Delta k_i V_s}
   \delta_{ij} \frac{75}{128128}  \nonumber \\
   & \times & \left[
   3003\left(P_{gg}(k_i)+\frac{1}{\bar{n}_g}\right)^2
   +1092 \left(P_{gg}(k_i)+\frac{1}{\bar{n}_g}\right) P_{gv}(k_i) + \frac{1652}{3} P_{gv}^2(k_i)  \right. \nonumber \\
   & + & \left. \frac{2}{15} P_{vv}(k_i)\left\{
   2065\left( P_{gg}(k_i)+\frac{1}{\bar{n}_g} \right) + 3126 P_{gv}(k_i)
   \right\}+\frac{1491}{17}P_{vv}^2(k_i)
   \right], \label{pgg:cov}  \\ 
   {\rm Cov}[\hat{P}_{gv}(k_i),\hat{P}_{gv}(k_j)] 
   & = &
   \frac{2(2\pi)^3}{2\pi k^2_i \Delta k_i V_s}
   \delta_{ij} \frac{105}{18304} \nonumber \\ 
   & \times & \left[ 3003\left(P_{gg}(k_i)+\frac{1}{\bar{n}_g}\right)^2 
   + 4420\left(P_{gg}(k_i)+\frac{1}{\bar{n}_g}\right)P_{gv}(k_i) + 2940P_{gv}^2(k_i)  \right. \nonumber \\
   & + & \left. 10 P_{vv}(k_i)\left\{
   147\left( P_{gg}(k_i)+\frac{1}{\bar{n}_g} \right)+226P_{gv}(k_i)
   \right\}+\frac{24185}{51}P_{vv}^2(k_i)
   \right], \label{pgv:cov2} \\ 
   {\rm Cov}[\hat{P}_{vv}(k_i),\hat{P}_{vv}(k_j)] 
   & = &
   \frac{2(2\pi)^3}{2\pi k^2_i \Delta k_i V_s}
   \delta_{ij} \frac{44100}{1537536} \nonumber \\
   & \times & \left[
   3003\left(P_{gg}(k_i)+\frac{1}{\bar{n}_g}\right)^2 
   + 1092\left(P_{gg}(k_i)+\frac{1}{\bar{n}_g}\right)P_{gv}(k_i) + \frac{23148}{5}P_{gv}^2(k_i) \right. \nonumber \\
   & + & \left. \frac{2}{5}
   P_{vv}(k_i)\left\{
   5787\left( P_{gg}(k_i)+\frac{1}{\bar{n}_g} \right) + 9810 P_{gv}(k_i)
   \right\}+\frac{14931}{17} P_{vv}^2(k_i)
   \right]. \label{pvv:cov} 
\end{eqnarray}
\end{widetext}

\section{Alternative parametrization of modified gravity}
\label{app:mg}

A useful modification to GR was proposed by \citep{amin}, which seeks 
to describe changes in the potential-density relationship
on large scales while leaving small scales unchanged from GR. 
The modification is expressed in the form of 
a power series in $a H/k$, which is the ratio of the proper scale of 
perturbations $a/k$ to the horizon size $1/H$. 
The Fourier-space analogue of the Poisson equation is assumed to be modified as follows
\begin{equation}
   -k^2 \Phi(a,{\bf k}) = 4 \pi a^2  G g(k) \bar{\rho} \, \delta(a,{\bf k}), 
    \label{eqn:mod_phi2}
\end{equation}
where $g(k) \equiv g_0(a) + g_1(a) \frac{a H}{k}$. 
The relation (\ref{eqn:mod_phi2}) converges to GR when $g_0=1$ 
and for scales much 
smaller than the horizon size,  i.e.  $a H/k \gg 1$. 
Note that we consider only the first two elements of the power
series. The linear term in the expansion (\ref{eqn:mod_phi2}) is 
characteristic for brane-world 
inspired models like DGP. 
The linear term is absent in scalar-tensor models including 
$f(R)$ models, where the first non-zero higher order term is 
quadratic in $a H/k$. 
For a thorough discussion of the parametrization and 
specific examples in different alternative gravity models
see \citep{amin}. 

Similar to the curvature potential $\Phi$, the Newtonian potential can be 
modified as \citep{amin}
\begin{equation}
   -k^2 \Psi(a,{\bf k}) = 4 \pi  a^2 G \mu(k) \bar{\rho} \, \delta(a,{\bf k}),
    \label{eqn:mod_psi2}
\end{equation}
where $\mu(k) \equiv \mu_0(a) + \mu_1(a) \frac{a H}{k}$.  
In GR $g_0=\mu_0=1$ and the scale dependence of both potentials vanishes. 
Thus, we parametrize the departure from GR using
$4$ parameters $g_0$, $g_1$, $\mu_0$ and $\mu_1$. They are used in the Fisher matrix analysis
with fiducial values are $g_0=\mu_0=1$ and $g_1=\mu_1=0$. 
This type of modification is supported by the fact that the growth of structures is sourced by the
Newtonian potential $\Psi$, so that $\mu_0$ and $\mu_1$ contain information 
about the effect of MG on the observed matter distribution. 
The growth equation takes the form \citep{JZ}
\begin{equation}
  \ddot{\delta}(a,k) +2 H(a) \dot{\delta}(a,k) + \mu(k) \frac{k^2}{a^2} \Phi(a,k) = 0. 
  \label{eqn:growth2}
\end{equation}

\begin{figure*}
\includegraphics[width=14cm]{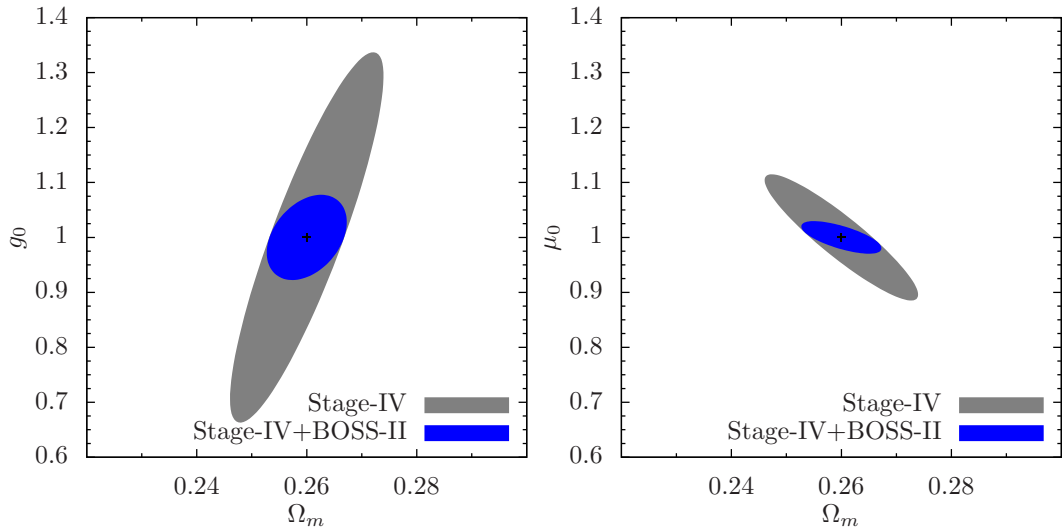}
\caption{
\label{des_boss} Achievable uncertainty on modified gravity parameters $g_0$, $\mu_0$ and matter density 
$\Omega_m$ for the future configuration of the Stage-IV imaging survey and the BOSS-II galaxy redshift survey.
Predictions for the Stage-IV survey only are shown in gray, for the Stage-IV and BOSS-II combined  -- in blue. 
There are shown $68\%$ confidence level contours. The dependence on the other parameters is marginalized out. 
}
\end{figure*}

The results for MG parameters with Stage-IV and BOSS-II surveys
are shown in Fig. \ref{des_boss}. 
The imaging Stage-IV survey (with the usual CMB prior) could constrain 
the scale-independent part of the effective gravitational constants 
as $\sigma(g_0) = 0.22$ and $\sigma(\mu_0) = 0.075$ 
(about a factor of $4$ smaller than those achievable with DES.)  
If we add information from BOSS-II spectroscopic survey we obtain 
$\sigma(g_0) = 0.050$ and $\sigma(\mu_0) = 0.018$ 
(about $2.5$ times smaller than for DES and BOSS-I. )

\begin{figure*}
\includegraphics[width=14cm]{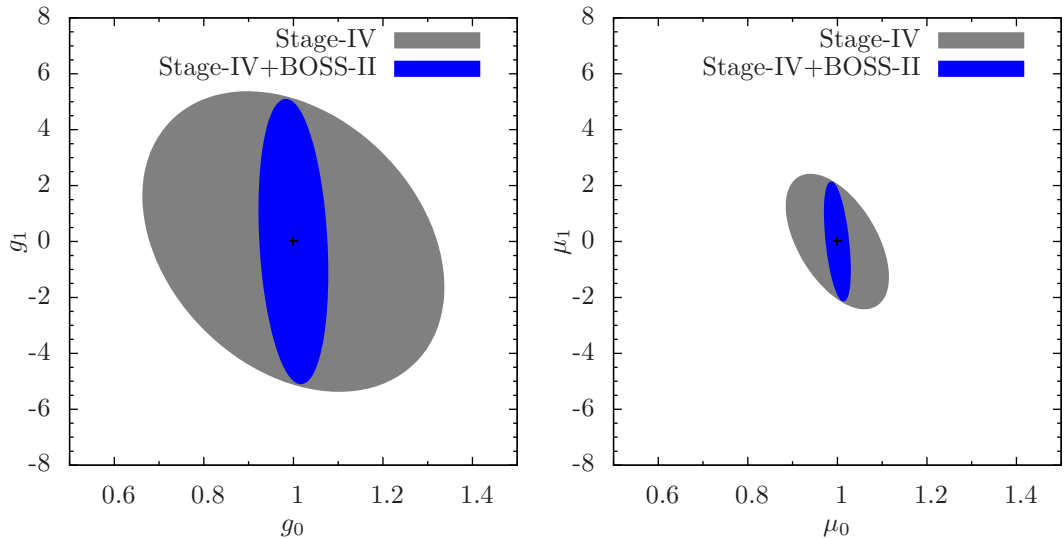}
\caption{
\label{des_boss1_kdep} Uncertainty in  modified gravity parameters $g_1$, $\mu_1$ which describe 
scale dependence of the growth of structure. There are shown $68\%$ confidence level contours 
for the Stage-IV alone and combined with BOSS-II. Note a lack of constrains on the scale dependence
of the modified gravity models for the assumed parametrization.
}
\end{figure*}

In Fig. \ref{des_boss1_kdep}  
we present joint constraints on the scale independent and 
dependent terms in the modification. 
The uncertainties on the scale dependent terms are much larger 
(compared to the constant terms): 
$\sigma(g_1) = 3.52$ and
$\sigma(\mu_1) = 1.39$. 
The weak constraints on the scale dependent terms 
relates to the fact that this dependence 
is important on large scales approaching the horizon,   
where the signal-to-noise is small. 
The integrated Sachs-Wolfe effect would be more promising 
to test for modifications on the largest scales. 

On the other hand, even constraining the scale independent modification 
lets us distinguish a class of brane-world models like DGP 
gravity from the standard LCDM one. 
The values of parameters under examination for the DGP model 
are $g_0=1.25$, $g_1=0.5$, $\mu_0=0.75$
$\mu_1=0.5$ at $z \sim 0.2$ (and closer to GR values for higher redshifts 
\citep{amin}). 
Thus the DGP model is several $\sigma$ away from the GR (see \cite{song07} for
current constraints on DGP). 
The differences in the growth history in these two gravity models 
helps discriminate them ($\mu_0$ is the most tightly constrained
parameter). 

We have also explored time evolving parametrization of $g$ and
$\eta$. With additional parameters, constraints generally get weaker,
but if one chooses a specific fiducial time evolution (such as linear 
or quadratic in $a$) it can become easier to test MG models. We leave 
a detailed 
exploration of time and scale dependence for future work. See
\cite{zhao09} for a recent study using principal components in the scale
and time dependence. 

\end{document}